# A microscopic nuclear collective rotation-vibration model: 2D submodel

P. Gulshani

NUTECH Services, 3313 Fenwick Crescent, Mississauga, Ontario, Canada L5L 5N1
Tel. #: 647-975-8233; matlap@bell.net

**Abstract**

We develop in this article a microscopic version of the successful phenomenological hydrodynamic Bohr-Davydov-Faessler-Greiner (*BDFG*) model for the collective rotation-vibration motion of a deformed nucleus. The model derivation is not limited to small oscillation amplitudes. The model generalizes the author's previous model to include interaction between collective oscillations in each pair of spatial directions, and to remove many of the previous-model approximations. To derive the model, the nuclear Schrodinger equation is canonically transformed to collective coordinates and then linearized using a constrained variational method. The associated transformation constraints are imposed on the wavefunction and not on the particle co-ordinates. This approach yields four self-consistent, time-reversal invariant, cranking-type Schrodinger equations for the rotation-vibration and intrinsic motions, and a self-consistency equation. To facilitate comparison with the *BDFG* model, simplify the solution of the equations, and gain physical insight, we restrict in this article the collective oscillations to only two space dimensions. For harmonic oscillator mean-field potentials, the equations are then solved in closed forms and applied to the ground-state rotational bands in some even-even light and rare-earth nuclei. The computed ground-state rotational band excitation energy, quadrupole moment and electric quadrupole transition probabilities are found to agree favourably with measured data and the results from mean-field, $Sp(3,R)$, and $SU(3)$ models.

## 1. Introduction

Experiments indicate that, at lower energies, the excited states of open-shell nuclei form various rotational bands and that these states have large electric quadrupole moments ($eQ_o$) and transition probability rates ($B(E2)$'s). Various methods and models have been developed over the years to describe, with various degrees of success, the rotational and vibrational motions implied by these empirical results. Among these methods and models are: (i) phenomenological *VMI* (Variable Moment of Inertia) model [1-6] and *BDFG* (Bohr-Davydov-Faessler-Greiner) model [6-10], (ii) group theoretic approaches [11,12], (ii) collective Hamiltonians derived from canonical transformations [13-26, refer to others in 27], (iii) diagonalization of the *A*-nucleon kinetic energy plus a phenomenological potential energy, with two to six fitting parameters including a pairing interaction in one case, in a suitably selected subspace of the harmonic oscillator irreducible representation of the non-compact group $Sp(3,R)$ [28-31], which generalizes the compact *SU*(3) approach, (iv) semi-classical time-dependent *HFB* (Hartree-Fock-Bogoliubov), constrained cranked *HFB*, and generator co-ordinate methods [32-34, and references therein], where the collective co-ordinates and Hamiltonian are identified with the expectation values of respectively multipole moment operators and microscopic nuclear Hamiltonian, and then quantized using Hamilton's principle and usual quantization rules, and (v) angular momentum projection methods [see references in [35]) using with microscopic interactions, with the goal of investigating automatic emergence of nuclear rotational spectrum [32,35,36].

Except for the phenomenological and semi-classical models, the models mentioned above have had various degrees of success in predicting the collective properties of light deformed nuclei. The phenomenological hydrodynamic rotation-vibration (deforming-rotating liquid-drop) model of *BDFG* [6-10], which uses four adjustable parameters for the two rotational and vibrational masses and the excitation energies of the first two excited states, has been especially successful for light and heavy nuclei. It is therefore of interest to understand the model from microscopic principles, and reveal its underlying assumptions and approximation and limitations, and hence have a more robust model. In other words, we would like to know how the collective model Hamiltonian and Schrodinger equation are related to the microscopic nuclear Hamiltonian and Schrodinger equation.

To achieve the above objective previous studies [13-26, refer to others in 27] canonically transformed the nuclear Hamiltonian to a set of collective Euler angles defining the orientation of the nuclear mass quadrupole moment tensor, a set of collective nuclear-shape co-ordinates chosen to be the principal-axes components of the quadrupole moment, and a set of intrinsic co-ordinates. The nucleon co-ordinates along the quadrupole-moment-tensor body-fixed axes were subjected to constraints, which partially defined the intrinsic co-ordinates. The transformation decomposed the Hamiltonian into a sum of three components: an intrinsic and a rotation-vibration components, and a component that coupled intrinsic and rotation-vibration systems. The rotational component turned out to be associated with an irrotational flow with the characteristically small irrotational-flow kinematic moment of inertia (the word "kinematic" refers to the moment inertia that excludes the contribution from the collective-intrinsic interaction). The coupling component, which involved the product of the components of the angular moment and intrinsic shear operators, would be expected to increase the irrotational-flow kinematic moment of inertia to that observed empirically (if a calculation could be performed). However, the transformation-related constraints imposed on the body-fixed particle co-ordinates, the unknown complicated intrinsic co-ordinates, and the complicated nature of the collective-intrinsic coupled component in the transformed Hamiltonian made it extremely difficult to perform any realistic calculation of the energy spectrum and nuclear properties.

In a previous publication [37,38], we circumvented the difficulties mentioned in the preceding paragraph by transforming the Schrodinger equation rather than the Hamiltonian, and by imposing the transformation-related constraints on the wavefunction rather than on the particle coordinates. In this way, we used the space-fixed particle co-ordinates at each stage of the transformation. Furthermore, the Euler angles orienting the rotating axes were chosen to describe a rigid flow with the resulting rigid-flow kinematic moment of inertia, in contrast to the Euler angles for irrotational flow chosen in most of the previous studies. In addition, the Coriolis interaction between the collective and intrinsic motions were eliminated by requiring the intrinsic wavefunction to have zero angular momentum (i.e., be spinless). In this way, the Coriolis interaction is shifted into the intrinsic system dynamics through the constraint of a spinless intrinsic wavefunction. The resulting rigid-flow rotor Schrodinger equation was then transformed to three collective vibration co-ordinates representing the three principle-axes components of the rigid-flow moment of inertia tensor using a collective-intrinsic product wavefunction for the nuclear wavefunction. The resulting partial differential equation for the coupled rotation-vibration-intrinsic motion was then reduced, using a constrained variational method, to three coupled ordinary differential equations for the collective rotation-oscillation motion, a usual equation for the intrinsic motion coupled algebraically to the collective motions, and a self-consistency equation. The nuclear potential energy was approximated by harmonic oscillator mean-field potentials with adjustable strength parameters for the collective and intrinsic systems.

To solve the resulting five governing equations, we used in [37,38] a number of approximations to simplify the equations and gain physical insight: (i) we ignored the constraints on the intrinsic system because earlier calculations seem to indicate that they may have relatively small effect, (ii) we assumed an axially symmetric collective motion, (iii) we neglected interaction between oscillations in each pair of spatial directions, and (iv) we replaced an anisotropic intrinsic operator by an isotropic operator. The resulting model [37,38]: (a) indicated that the interaction between the collective and intrinsic motions reduced the rigid-flow kinematic moment of inertia in the first excited state and increased it in the higher excited states, (b) indicated that the ground-state rotational band terminated at some low value of the angular momentum when the self-consistency among the governing model equations was violated because of centrifugal stretching, (c) predicted reasonably well the excitation energy of the members of the ground-state rotational band in light and rare-earth nuclei, (d) predicted that the $eQ_o$ and the $B(E2)$'s increased monotonically with the angular momentum quantum number $J$. The prediction in item (d) is not consistent with empirical results and the prediction of the $SU(3)$ [39] and $Sp(3,R)$ [28,29] models, which indicate that $eQ_o$ decreases with $J$, and $B(E2)$ increases at low values of $J$ and decreases at higher values of $J$.

In this article, the approximations in the previous study [37,38], referred to in items (ii) to (iv) the preceding paragraph, are removed to obtain the observed behaviour in $eQ_o$ and $B(E2)$ with $J$. To facilitate comparison with the $BDFG$ model, simplify the solution of the equations, and gain physical insight, we restrict in this article the collective oscillations to only two space dimensions. This restriction allows us to avoid dealing with complications (of mathematical and physical-interpretation nature, such as wobbling motion, associated with using three vibration quantum numbers) arising from a full 3-D vibration model. This 2-D submodel of the microscopic collective rotation-vibration-intrinsic model is derived in Section 2. In section 3, we solve in closed forms the model governing equations, obtain expressions for the model coupling parameters and the excitation energy for the members of the rotational bands, and solve the algebraic equations for the coupling parameters. In Section 4, we discuss how the model coupling parameters affect the nuclear energy, deformation, and collective and intrinsic motions, and generate self-consistency and feedback mechanisms among these motions. In Section 5, we compare the model with *MVI*, *BDFG*, and mean-field models including the cranking models. In Section 6, we use the model to predict the excitation energy, moment of inertia, and quadrupole moment, and transition probability rates for the members of the ground-state rotational band in light and rare-earth nuclei. The article is concluded in Section 7.

## 2. Derivation of microscopic collective-intrinsic model

The model is derived by transforming the nuclear Schrodinger equation to collective Euler angles and vibration co-ordinates with constraints imposed on the wavefunction. The transformation is performed in two steps.

In the first step, we use the microscopic rotational model derived in [40,41] and described briefly here. We use the rotational-model product wavefunction:

$$\Psi_{J\mathcal{M}} = \sum_{K=-J}^{J} \mathcal{D}_{\mathcal{M}K}^{J}(\theta_s) \cdot \Phi_{JK}(x_{ni}), \quad \text{where} \quad \hat{J}_A \Phi_{JK}(x_{ni}) = 0, \quad \frac{\partial \theta_s}{\partial x_{nj}} = \sum_{k=1}^{3} \chi_{jk}^{s} x_{nk} \quad (1)$$

where $\mathcal{D}_{\mathcal{M}K}^{J}$ is the Wigner rotation matrix, $\hat{J}_A$ is the $A^{\text{th}}$ component of the total angular momentum operator, $\chi^s$ ($s = 1,2,3$) are three non-zero, arbitrary anti-symmetric 3x3 matrices, and $x_{nj}$ ($n = 1,...,A; \; j = 1,2,3,$ where $A =$ nuclear mass number) are the space-fixed nucleon co-ordinates. Transforming the nuclear Schrodinger equation to the co-ordinates $\theta_s$ and its conjugate angular momentum $\hat{J}_s$, we obtain [41]:

$$\left( -\frac{\hbar^2}{2M} \sum_{n,j=1}^{A,3} \frac{\partial^2}{\partial x_{nj}^2} + \hat{V} - E \right) \Phi_{JK'} + \frac{1}{2} \sum_{\mathcal{M},K=-J}^{+J} \sum_{A,B=1}^{3} \Phi_{JK} \mathfrak{I}_{AB}^{rig^{-1}} \mathcal{D}_{\mathcal{M}K'}^{J*} \hat{J}_B \hat{J}_A \mathcal{D}_{\mathcal{M}K}^{J} = 0 \quad (2)$$

where $M$ is the nucleon mass, $\hat{V}$ is a rotationally invariant two-body interaction, $\hat{J}_A$ is the $A^{\text{th}}$ component of the total angular momentum operator along the $A^{\text{th}}$ co-ordinate axis defined by $\theta_s$, and $\mathfrak{I}_{AB}^{rig^{-1}}$ is the component of the inverse of the rigid-flow moment of inertia tensor defined by:

$$\mathfrak{I}^{rig} \equiv Tr(Q) - Q, \quad \text{and} \quad Q_{jk} \equiv M \sum_{n=1}^{A} x_{nj} x_{nk} \quad (3)$$

where $Q$ the mass quadrupole-moment tensor. The rigid-flow kinematic moment of inertia $\mathfrak{I}_{AB}^{rig}$ appears in Eq. (2) because of the choice of $\theta_s$ and the spinless intrinsic state given in Eq. (1). These choices also cause the Coriolis interaction term to vanish in the transformed Schrodinger equation Eq. (2). In this way, the Coriolis interaction has been shifted into the intrinsic system dynamics through the constraint of a spinless intrinsic wavefunction. A related consequence of the choices in Eq. (1) is that the nucleon kinematic velocity field has the rigid-flow character [41]. These results differ from the irrotational-flow moment of inertia and two-component (rigid plus irrotational) flow velocity fields in other studies [13-16,18-26, refer to others in 27], where the intrinsic state carries some angular momentum, and the resulting non-zero Coriolis interaction is determined by a shear (rotating-deforming) operator. The two models would of course yield the same final results if the governing equations could be solved. Specifically, in our model, the interaction of the intrinsic motion with the oscillations coupled to the centrifugal stretching reduces the rigid-flow kinematic moment of inertia (as shown in Section 5) to empirically observed values, and in [13-16,18-26, refer to others in 27] studies the shear operator would be expected to increase the irrotational-flow kinematic moment of inertia.

In Eq. (2), we evaluate the action of angular momentum operators on $\mathcal{D}^{J}_{\mathcal{M}K}$ and then integrate over $\theta_s$ to obtain the effective rotation-intrinsic asymmetric rotor Schrodinger equation [41]:

$$\begin{aligned}
E\Phi_{JK} = &\left\{-\frac{\hbar^2}{2M}\sum_{n,j}\frac{\partial^2}{\partial x_{nj}^2}+\hat{V}+\frac{\hbar^2}{4}\left(\mathfrak{I}_{11}^{rig^{-1}}+\mathfrak{I}_{22}^{rig^{-1}}\right)\left[J(J+1)-K^2\right]+\frac{\hbar^2 K^2}{2}\mathfrak{I}_{33}^{rig^{-1}}\right\}\Phi_{JK} \\
&+\frac{\hbar^2}{8}\left(\mathfrak{I}_{11}^{rig^{-1}}-\mathfrak{I}_{22}^{rig^{-1}}-2i\mathfrak{I}_{12}^{rig^{-1}}\right)\sqrt{(J+K+2)(J-K-1)(J+K+1)(J-K)}\,\Phi_{JK+2} \\
&+\frac{\hbar^2}{8}\left(\mathfrak{I}_{11}^{rig^{-1}}-\mathfrak{I}_{22}^{rig^{-1}}+2i\mathfrak{I}_{12}^{rig^{-1}}\right)\sqrt{(J-K+2)(J+K-1)(J-K+1)(J+K)}\,\Phi_{JK-2} \quad (4) \\
&+\frac{\hbar^2}{4}(2K-1)\left(\mathfrak{I}_{13}^{rig^{-1}}-i\mathfrak{I}_{23}^{rig^{-1}}\right)\sqrt{(J+K+1)(J-K)}\,\Phi_{JK+1} \\
&+\frac{\hbar^2}{4}(2K+1)\left(\mathfrak{I}_{13}^{rig^{-1}}+i\mathfrak{I}_{23}^{rig^{-1}}\right)\sqrt{(J-K+1)(J+K)}\,\Phi_{JK-1}
\end{aligned}$$

In this article, we ignore the terms in Eq. (2) that involve the relatively small $(\mathfrak{I}_{11}^{rig^{-1}}-\mathfrak{I}_{22}^{rig^{-1}})$ (non-axial terms) and small off-diagonal elements $\mathfrak{I}_{AB}^{rig^{-1}}$ (or equivalently $Q_{AB}$) $(A\neq B)$ (in fact their averages generally vanish) as may be inferred from the characteristic $J(J+1)$ rotational energy spectra of even-even nuclei. Eq. (4) then reduces to:

$$\left\{-\frac{\hbar^2}{2M}\sum_{n,j}\frac{\partial^2}{\partial x_{nj}^2}+\hat{V}+\frac{\hbar^2}{2\mathfrak{I}_{11}^{rig}}\left[J(J+1)-K^2\right]+\frac{\hbar^2 K^2}{2\mathfrak{I}_{33}^{rig}}\right\}\Phi_{JK}=E\Phi_{JK} \quad (5)$$

In the second step of the transformation, we transform Eq. (5) to the collective vibration (rigid-flow-moment-of inertia) co-ordinates defined by:

$$R_1 \equiv \frac{\mathfrak{I}_{11}^{rig}}{M}=\sum_{n=1}^{A}(y_n'^2+z_n'^2)\,,\quad R_2 \equiv \frac{\mathfrak{I}_{22}^{rig}}{M}=\sum_{n=1}^{A}(x_n'^2+z_n'^2)\,,\quad R_3 \equiv \frac{\mathfrak{I}_{33}^{rig}}{M}=\sum_{n=1}^{A}(x_n'^2+y_n'^2) \quad (6)$$

(where the prime superscript indicates the co-ordinates along the body-fixed axes[1] given by the transformation $x_{nj} = \sum_{A=1}^{3} R_{Aj}(\theta_s) x'_{nA}$ where $R_{Aj}$ are the elements of an orthogonal matrix). In transforming Eq. (5) to the co-ordinates $R_1$, $R_2$, and $R_3$ in Eq. (6), we use the product wavefunction:

$$\Phi_{JK} = F_1(R_1) \cdot F_2(R_2) \cdot F_3(R_3) \cdot \phi_{JK}(x_{nk}) \tag{7}$$

where the spherically symmetric[2] intrinsic (such as shell-model or *HFB*) wavefunction $\phi_{JK}$ is subject to the constraints:

$$\frac{\partial}{\partial R_1} \phi_{JK} = 0, \quad \frac{\partial}{\partial R_2} \phi_{JK} = 0, \quad \frac{\partial}{\partial R_3} \phi_{JK} = 0 \tag{8}$$

We then obtain the result:

$$\sum_{n=1}^{A} \frac{\partial^2}{\partial x_n^2} \Phi_{JK} = 2A(F_1 F_3 \frac{dF_2}{dR_2} + F_1 F_2 \frac{dF_3}{dR_3}) \cdot \phi_{JK}$$
$$+ \sum_{n=1}^{A} x_n^2 (4 F_1 F_3 \frac{d^2 F_2}{dR_2^2} + 8 F_1 \frac{dF_2}{dR_2} \cdot \frac{dF_3}{dR_3} + 4 F_1 F_2 \frac{d^2 F_3}{dR_3^2}) \cdot \phi_{JK} \tag{9}$$
$$+ 4(F_1 F_3 \frac{dF_2}{dR_2} + F_1 F_2 \frac{dF_3}{dR_3}) \cdot \sum_{n=1}^{A} x_n \frac{\partial}{\partial x_n} \phi_{JK} + F_1 F_2 F_3 \cdot \sum_{n=1}^{A} \frac{\partial^2}{\partial x_n^2} \phi_{JK}$$

and similar expressions for the derivatives of $\Phi_{JK}$ with respect to $x_{n2}$ and $x_{n3}$. To derive Eq. (9), we have used the approximation of replacing $x'_{nA}$ by $x_{nk}$. We obtain the same results if we neglected the small off-diagonal elements of $Q_{AB}$ ($A \neq B$) and chose the non-zero anti-symmetric matrix $\chi^s$ in Eq. (1) to satisfy the condition:

$$\chi^s_{AB} \omega^s_{BA} C'_{BA} = -\chi^s_{AC} \omega^s_{CA} C'_{CA} \tag{10}$$

where *A*, *B*, *C* = 1, 2 ,3 are in cyclic order, and

$$\omega^s_{AB} \equiv \sum_{j=1}^{3} \frac{\partial R_{Aj}}{\partial \theta_s} R_{Bj}, \quad C'_{BC} \equiv \sum_{n=1}^{A} (x'^2_{nB} - x'^2_{nC}) \tag{11}$$

---

[1] $R_1$, $R_2$, and $R_3$ can be and are chosen independent collective co-ordinates because *A*-3 of the particle co-ordinates $x'_{nA}$ are independent of each other.

[2] The prescriptions in Eqs. (7) and (8) extract the quadrupole-moment distribution part of the wavefunction $\Phi_{JK}$ leaving the intrinsic wavefunction $\phi_{JK}$ spherically symmetric, noting that higher order deformations are found empirically to be relatively small [7,42] and hence are usually not considered at not too high angular momenta.

For $\hat{V}$ in Eq. (5) we use the harmonic-oscillator mean-field potential:

$$\hat{V} = \frac{M\omega_{\text{int}}^2}{2}\sum_{n=1}^{A} r_n^2 + \frac{M\omega_{v1}^2}{2} R_1 + \frac{M\omega_{v2}^2}{2} R_2 + \frac{M\omega_{v3}^2}{2} R_3 \qquad (12)$$

where $r_n^2 \equiv x_n^2 + y_n^2 + z_n^2$, and the first term in Eq. (12) is the restoring potential for the intrinsic system, and the remaining terms are the restoring potentials for the oscillations in the collective rigid-flow moment of inertia variables $R_1$, $R_2$ and $R_3$ [3].

Substituting Eqs. (7), (8), (9) and (12) into Eq. (5), we obtain a partial differential equation for the transformed Schrodinger equation where the collective oscillations in each of the three spatial directions are functionally coupled to each other and to rotation and intrinsic motions. To reduce this coupling to an algebraic coupling effectively linearizing the equation, we apply to the transformed Schrodinger equation a constrained variational method, as done previously [43,44]. To achieve this goal, we take the expectation of the equation and we vary separately each of the wavefunctions $F_1^*$, $F_2^*$, $F_3^*$, and $\phi_{JK}^*$ (i.e., we use the Rayleigh-Ritz variational method) subject to the normalization and energy minimization conditions: $\langle F_1|F_1\rangle = \langle F_2|F_2\rangle = \langle F_3|F_3\rangle = \langle \phi_{JK}|\phi_{JK}\rangle = 1$, $\partial E/\partial F_1^* = \partial E/\partial F_2^* = \partial E/\partial F_3^* = \partial E/\partial \phi_{JK}^* = 0$, for arbitrary variations in $F_1^*, F_2^*, F_3^*$, and $\phi_{JK}^*$. The variation yields four coupled, self-consistent, time-reversal invariant, cranking-type ordinary differential equations, and a fifth self-consistency equation. In this article, we simplify the solutions of equations to gain physical insight and to facilitate comparison with the *BDFG* model, and we retain only the equations that are related to the collective and intrinsic motions in the direction 1 and 3. This restriction appears to be a reasonable simplifying approximation for the ground-state rotational band of nuclei, with which we are concerned in this article, and because doing so avoids added complexity in mathematics and their physical interpretation (such as wobbling motion associated with a third vibration quantum number in a full 3-D vibration model). We then obtain the following three self-consistent, time-reversal invariant cranking-type ordinary differential Schrodinger equations and a self-consistency equation:

$$\left[ R_1^2 \frac{d^2}{dR_1^2} + R_1\left(\bar{a}_1 - \frac{1}{4}\beta_3 R_1\right)\frac{d}{dR_1} - \frac{J(J+1)-K^2}{8} + \frac{1}{4}\varepsilon_1 R_1 - \frac{b_{v1}^2}{4} R_1^2 \right]|F_1\rangle = 0 \qquad (13)$$

$$\left[ R_3^2 \frac{d^2}{dR_3^2} + R_3\left(\bar{a}_3 - \frac{1}{4}\beta_1 R_3\right)\frac{d}{dR_3} - \frac{K^2}{4} + \frac{1}{4}\varepsilon_3 R_3 - \frac{b_{v3}^2}{4} R_3^2 \right]|F_3\rangle = 0 \qquad (14)$$

---

[3] To satisfy the constraint in Eqs. (8), we also need to add a two-body (such as separable monopole-monopole) interactions to the right-hand-side of Eq. (12) as we did in [43,44]. In this article, we do not do so for simplicity and to gain physical insight into the results and because, using the method in [43,44], it can be shown that the effects of the constraints in Eqs. (8) are relatively small. We also ignore, in this article, the zero-angular momentum constraint in Eq. (1) since calculations [43,44] seem to indicate that it may have a relatively small effect.

$$\left(-\sum_{n=1}^{A}\nabla_n^2 + \beta_1 \cdot \tilde{B}_3 + \beta_3 \cdot \tilde{B}_1 + b_{int}^2 \sum_{n=1}^{A} r_n^2 - \varepsilon_{int}\right)|\phi_{JK}\rangle = 0 \qquad (15)$$

$$a_3 \cdot \beta_1 + a_1 \cdot \beta_3 = \varepsilon_s \qquad (16)$$

where:

$$\bar{a}_1 \equiv a_3 - \frac{1}{4}\gamma_3, \quad \bar{a}_3 \equiv a_1 - \frac{1}{4}\gamma_1, \quad a_k \equiv \langle \phi_{JK}|\tilde{B}_k|\phi_{JK}\rangle, \quad \gamma_k \equiv -4\langle F_k|R_k \frac{d}{dR_k}|F_k\rangle, \quad k=1,3 \qquad (17)$$

$$\beta_k \equiv -4\langle F_k|\frac{d}{dR_k}|F_k\rangle, \quad \tilde{B}_k \equiv \frac{1}{2}\sum_{n=1}^{A}\left(x_{nk}\frac{\partial}{\partial x_{nk}} + \frac{\partial}{\partial x_{nk}}x_{nk}\right) \qquad (18)$$

and $\varepsilon_1$, $\varepsilon_3$, $\varepsilon_{int}$, and $\varepsilon_s$ are functions of the reduced energy $\varepsilon \equiv 2ME/\hbar^2$ and the system parameters in Eqs. (17) and (18). We note that the coupling self-consistency parameters appearing in Eqs. (13) to (16) are expectation values of collective vibration displacement and intrinsic compression-dilation operators.

## 3. Solution of eqs (13)-(16)

We readily obtain, from the literature as discussed in [43,44], the solutions of Eqs. (13)-(15) in closed forms (refer to footnote 3). In particular, Eq. (15) is solved exactly in Cartesian co-ordinate system avoiding the approximate solution obtained in [37,38] using spherical co-ordinate system. The solutions of Eqs. (13)-(15) are used to evaluate then coupling parameters in Eqs. (17) and (18) and $\varepsilon_1$, $\varepsilon_3$, $\varepsilon_{int}$, and $\varepsilon_s$[4] in Eqs. (13)-(16). In particular, we obtain the reduced excitation energy:

$$\varepsilon = 2b_1 \Sigma_1 + 2b_3 \Sigma_3 + 2\bar{b}_{v1}(2n_1 + \bar{a}_1 + 2k_1) + 2\bar{b}_{v3}(2n_3 + \bar{a}_3 + 2k_3) \\ - \frac{1}{2}\beta_3 \bar{a}_1 - \frac{1}{2}\beta_1 \bar{a}_3 - \beta_1 a_3 - \beta_3 a_1 + \frac{1}{4}\beta_1 \gamma_3 + \frac{1}{4}\beta_3 \gamma_1 \qquad (19)$$

where:

$$2k_1 \equiv -(\bar{a}_1 - 1) + \sqrt{(\bar{a}_1 - 1)^2 + [J(J+1) - K^2]/2}, \quad 2k_3 \equiv -(\bar{a}_3 - 1) + \sqrt{(\bar{a}_3 - 1)^2 + K^2}, \quad \bar{a}_1, \bar{a}_3 > 1 \qquad (20)$$

$$b_1 \equiv \sqrt{b_{int}^2 + \beta_3^2/4}, \quad b_3 \equiv \sqrt{b_{int}^2 + \beta_1^2/4}, \quad a_1 \equiv \frac{\beta_3 \Sigma_1}{2b_1}, \quad a_3 \equiv \frac{\beta_1 \Sigma_3}{2b_3}, \quad \Sigma_k \equiv \sum_{m_k=0}^{m_{kf}}(m_k + 1/2) \qquad (21)$$

$$\bar{a}_1 = a_3 - \frac{1}{2}\bar{a}_3 + \frac{\beta_1}{8} \cdot \frac{\bar{a}_3 + 2k_3}{\bar{b}_{v3}}, \quad \bar{a}_3 = a_1 - \frac{1}{2}\bar{a}_1 + \frac{\beta_3}{8} \cdot \frac{\bar{a}_1 + 2k_1}{\bar{b}_{v1}} \qquad (22)$$

---

[4] The solutions of Eqs. (13) and (14) given in this article differ from those of Faessler-Greiner rotation-vibration model [9,10] in three respects: (i) our solutions are not limited to small amplitude oscillations about a mean deformation, (ii) the kinematic moment of inertia is not an adjustable parameter but rather is a dynamical variable (specifically is the rigid-flow moment), (iii) Eqs. (13) and (14) include the interaction between the two collective oscillations, and (iv) Eqs. (13)-(15) include the interaction between rotation-vibration and intrinsic motions.

$$\beta_1 = \frac{8\bar{b}_{v1}(\bar{a}_1-1)}{3(\bar{a}_1+2k_1-1)} - \frac{4\bar{b}_{v3}(\bar{a}_3-1)}{3(\bar{a}_3+2k_3-1)}, \quad \beta_3 = \frac{8\bar{b}_{v3}(\bar{a}_3-1)}{3(\bar{a}_3+2k_3-1)} - \frac{4\bar{b}_{v1}(\bar{a}_1-1)}{3(\bar{a}_1+2k_1-1)} \quad (23)$$

$$\bar{b}_{v1}^2 = \frac{1}{16}\beta_3^2 + b_{v1}^2, \quad \bar{b}_{v3}^2 = \frac{1}{16}\beta_1^2 + b_{v3}^2 \quad (24)$$

$n_1, n_3 = 0,1,2,3,\ldots\infty$ are the quantum numbers for the collective oscillations in respectively 1 and 3 spatial directions (and may be identified with the so-called beta and gamma band heads), $\Sigma_k$ is the total oscillator particle-occupation number in $k^{\text{th}}$ direction, $m_k$ is a harmonic oscillator quantum number, and $m_{kf}$ is the value of $m_k$ for the last particle-occupied (Fermi) level.

To obtain the solution to Eqs. (20)-(24), we first solve Eqs. (23) and (24) for $\bar{b}_{v1}^2$, $\bar{b}_{v3}^2$, $\beta_1$, and $\beta_3$ in terms of $b_{v1}^2$, $b_{v3}^2$, $\bar{a}_1$, $\bar{a}_3$, $2k_1$, and $2k_3$. We then solve iteratively Eqs. (22) for $\bar{a}_1$ and $\bar{a}_3$ and hence for $a_1$, $a_3$, $b_1$, $b_3$, and $\varepsilon$ in terms of $b_{v1}^2$, $b_{v3}^2$, $b_{int}^2$, $\Sigma_k$, $n_1, n_3$, $J$, and $K$. The solutions also yield the value of the rotational-band cutoff angular momentum $J_c$, which occurs when the value of $\bar{a}_1$ is near unity, and hence violating the condition on $\bar{a}_1$ in Eq. (20).

The excitation energy of a member of the ground-state rotational band is defined by:

$$\Delta E_J \equiv \frac{\hbar^2}{2M}\left[\varepsilon(J) - \varepsilon(J=0)\right] \quad (25)$$

where $\varepsilon$ is given in Eq. (19).

Generally, the moment of inertia $\mathfrak{I}_J$ for a given member of a rotational band with angular momentum $J$ is defined by [45]:

$$\frac{2\mathfrak{I}_J}{\hbar^2} = \frac{4J-2}{\Delta E_J - \Delta E_{J-2}} \quad (MeV)^{-1} \quad (26)$$

where $\Delta E_J$ is either the predicted or measured excitation energy.

For the application of the model to the ground-state rotational band in nuclei, we set $K=0$, $2k_3=0$, $n_1=0$, and $n_3=0$ in the above equations.

## 4. Discussion of model feedback mechanisms and their impacts

It may be evident that the intrinsic coupling parameters $\bar{a}_1$ and $\bar{a}_3$ appearing in the collective oscillations governing Eqs. (13) and (14) enhance the collective vibration displacements in the directions 1 and 3 respectively. Similarly, the collective vibration parameters $\beta_1$ and $\beta_3$ in Eqs. (13) and (14) retard the vibration displacements in the directions 1 and 3 respectively. In the intrinsic Eq. (15), the parameters $\beta_1$ and $\beta_3$ generate compression-dilation motion in the intrinsic system. Each of these motions is affected by the rotational motion or excitation through the centrifugal terms $J(J+1)$ and $K^2$ in Eqs. (13) and (14) or in the parameters $2k_1$ and $2k_3$ in Eq. (20) since $\bar{a}_1, \bar{a}_3, \beta_1$, and $\beta_3$ in Eqs. (22) and (23) depend on $2k_1$ and $2k_3$ and hence $J$ and $K^2$. Eqs. (17) and (18) for the coupling parameters indicate that these motions feed back on each other, creating a self-consistency among Eqs. (13)-(16). This self-consistency is mediated by the coupling parameters $\bar{a}_1$, $\bar{a}_3$, $\beta_1$, $\beta_3$, $a_1$, and $a_3$. Eqs. (22) arise from this self-consistency. The interactions and their feedbacks discussed above couple strongly the rotation, vibration, and intrinsic motions, and any separation of the energies associated with these motions may not be possible even at the lowest energy or lowest $J$ value. This may become more evident when we expand the parameters and energy in powers of $J(J+1)$.

From the feedback mechanisms discussed above, it is not surprising that the excitation energy in Eq. (19) explicitly contains all the coupling parameters $\bar{a}_1, \bar{a}_3, \beta_1, \beta_3, \gamma_1$, and $\gamma_3$, and hence every term in Eq. (19) is affected by all the other terms. In Eq. (19), the first two terms are the intrinsic energies. The third and fourth terms are the energy eigenvalues for the collective oscillations (including the effects of centrifugal stretching $J(J+1)$ and $K^2$ in the $2k_1$ and $2k_3$ parameters in Eq. (20) in respectively 1 and 3 spatial directions). The fifth to eighth terms are the energies arising from the interaction between the collective-vibration displacement and intrinsic dilation-compression (i.e., related to the second terms in Eqs. (13) and (14)). The ninth and tenth terms are purely the energies arising from the interaction between the collective vibrations in the two spatial directions 1 and 3. The fifth to tenth terms in Eq. (19) are responsible for reducing the value of the rigid-flow kinematic moment of inertia to the measured value without using any pairing interaction in the intrinsic system.

Calculation presented in Section 6 shows that $\bar{a}_1$ and $\beta_1$ decrease and $\bar{a}_3$ and $\beta_3$ increase with $J$. It follows from Eqs. (21) and (24) that the effective intrinsic and collective oscillation mean-field strengths $b_1$ and $\bar{b}_{v1}$ increase, and $b_3$ and $\bar{b}_{v3}$ decrease with $J$. Therefore, rotational excitation increases the effective intrinsic mean-field strength in the direction 1 and reduces it in the direction 3, thereby changing the intrinsic energy $2b_1\Sigma_1 + 2b_3\Sigma_3$ in Eq. (19). Similarly, the rotational-excitation induced increase in $\bar{b}_{v1}$ and reduction in $\bar{b}_{v3}$ respectively reduces the expectation of $R_1$ and increases the expectation of $R_3$ causing the quadrupole moment to generally decrease with $J$.

## 5. Comparison of models

This section presents a comparison of our collective and other such models.

For large values of $\bar{a}_1$ and $\bar{a}_3$, we may expand the energies in Eqs. (19) and (25) in powers of $J(J+1)$ to obtain the well-known phenomenological Variable-Moment-of-Inertia model expression [1-6]:

$$\Delta E_J \equiv \frac{\hbar^2}{2M}\left[\varepsilon(J) - \varepsilon(J=0)\right] = AJ(J+1) - BJ^2(J+1)^2 + CJ^3(J+1)^3 + \ldots$$
$$\equiv \frac{\hbar^2 J(J+1)}{2\mathfrak{J}_J} \tag{27}$$

where the coefficients $A$, $B$, $C$, etc. and the moment of inertia $\mathfrak{J}_J$ are functions of $b_{int}$, $b_{v1}$, $b_{v3}$, $\Sigma$, and $J$. An expansion resembling that in Eq. (27) was also obtained in the Faessler-Greiner rotation-vibration model after including the vibration-rotation interaction (a second-order term in their expansion in the deformation parameters) [10].

The Faessler-Greiner rotation-vibration model [10] describes an incompressible, irrotational-flow[5] rotation and small-amplitude oscillations of a tri-axial nucleus about an axially-symmetric deformed equilibrium shape. The nucleus is represented by a liquid-drop with small-amplitude waves travelling on its surface. Its Hamiltonian is expanded in quadratic terms in wave amplitudes and velocities, excluding cross terms associated with the interaction among the oscillations in the three spatial directions. The Hamiltonian is transformed to a rotating frame and to the liquid-drop shape parameters and then it is quantized. In this Hamiltonian, vibration and rotation masses and the potential energy strengths are taken to be adjustable parameters. In our collective model, the rotational kinetic energy term in Eq. (4) is identical to that in the Faessler-Greiner rotation-vibration model [10], except for the terms containing the off-diagonal elements $\mathfrak{J}_{AB}^{rig^{-1}}$ ($A \neq B$) and for the rigid-flow moment of inertia $\mathfrak{J}_{AA}^{rig^{-1}}$. When a transformation to the co-ordinate $\rho = \sqrt{R}$ is made, Eqs. (13) and (14) become similar to those in the Faessler-Greiner rotation-vibration model [10], except for the second term in the square brackets in Eqs. (13) and (14) arising from the interaction between the oscillations in 1 and 3 directions (expressed by the terms in $\beta_1$, and $\beta_3$), and between the oscillation-rotation and the intrinsic motion (expressed by the terms in $\bar{a}_1$ and $\bar{a}_3$)(refer also to footnote 4). Because of these differences between the Faessler-Greiner rotation-vibration model [10] and the model presented in this article, the expression for the excited-state energy $\varepsilon$ in Eq. (19) differs from that in the Faessler-Greiner model. Furthermore, even the third and fourth terms in Eq. (19), which are the oscillation-rotation energy eigenvalues (where the angular momentum appears under the square-root sign in $2k_1$ and $2k_3$ in Eq. (20)), differ from that in the Faessler-Greiner model because we do we do not limit the analysis to small amplitude oscillations.

The cutoff angular momentum $J_c$ feature predicted by the model is shared with that of the $SU(3)$ [5,46] and cranking [5] models. In our model as in the other models, $J_c$ increases with the mass number. Therefore, high angular-momentum phenomena (such as backbending) in heavy nuclei could occur in the model before a rotational band terminates.

---

[5] However, the constant factors in the rotation and vibration masses used in the Faessler-Greiner model are not those for irrotational flow but are rather fitted to the measured excitation energies of the first excited $2^+$ states. The impact of this course of action and its possible inconsistency need to be studied because the measured inertia masses do not have irrotational-flow character. The kinematic moment of inertia in the Faessler-Greiner model is proportional to the square of the deformation parameter, whereas the rigid-flow kinematic moment of inertia in our model is insensitive to very small value of the deformation.

The model in this article offers an alternative physical mechanisms for gradual and large increases in the moment of inertia and for a band termination to those provided by other models. Nevertheless, the mechanisms in this and other models closely parallel one another as the following description attempts to show.

In the successful microscopic mean-field approaches (such as time-dependent *HFB*, constrained cranked *HFB*, generator co-ordinate, and projected *HFB* methods [32-34, 47, and references therein]) for describing nuclear vibration-rotation motion, the Coriolis interaction is a dominant force. These models predict [32,48, and references therein] that the experimentally observed gradual increase with $J$ in the moment of inertia, and hence the resulting deviation of the rotational energy spectrum from $J(J+1)$ rule for a strongly-bound system, is primarily due to a Coriolis-force-induced weakening of nuclear pairing correlation and a gradual alignment of the axes of many quasi-particle orbits along the rotation axis with increasing $J$ (known as Coriolis anti-pairing or *CAP* effect) to achieve the lowest possible system energy [47,49]. Centrifugal stretching and the resulting increase in deformation play a minor role in this deviation. At a sufficiently high $J$, the *CAP* effect breaks up a nucleon pair in the highest angular momentum orbitals and align their spins along the rotation axis (known as the rotation alignment or *RAL* effect), causing a sudden increase in the moment of inertia (known as back-bending effect). But the moment of inertia remains less than the rigid-flow (or rigid-body) value consistent with the measured value of the moment of inertia. Eventually, *RAL* causes the disappearance or quenching of the pairing correlation, and hence a transition from a super-fluid to a normal-fluid nuclear state, and a corresponding increase in the moment of inertia. However, the moment of inertia still remains less than the rigid-flow value (due to the shell structure and internal counter currents [50], which the cranking model seems to predict [51,52 and references therein]) consistent with the measured moment. For nuclei with moderate deformations in this fully aligned state with the maximum angular momentum, the rotation is not collective but rather has an independent particle nature. In this state, the collective ground-state rotational band therefore terminates, and the nucleus acquires an oblate particle distribution about the rotation axis, which therefore becomes the symmetry axis as well [48,53,54], similarly to a classical (macroscopic) object. In this non-collective state, the collective nuclear deformation therefore disappears and $E2$ (electric quadrupole moment) transitions vanish or become small [5,55,56]. Excited-state beta and gamma rotational bands (including states with triaxial shapes) would have different band-termination $J$ values and would require a separate consideration. For nuclei with large deformations, the nuclear spin remains less than the maximum as the nucleus continues to deform (due to the centrifugal stretching) towards necking and fission with increasing angular momentum [54].

In the model derived in this article, the Coriolis interaction in the nuclear Schrodinger Eqs. (2), (4), (5), and (13)-(15), which couples the rotational and non-rotational degrees of freedom, has been transformed away and eliminated completely by the choice in Eq. (1) of the rotating frame. In this way, the Coriolis interaction has been shifted into the intrinsic system dynamics through the constraint of a spinless intrinsic wavefunction. In the current model, this state is approximated by a state where each pair of nucleons is placed in a pair of time-reversed deformed Nilsson's orbitals, or in other words, by having a state of paired nucleon with approximately zero angular momentum. The elimination of the Coriolis interaction results in: (i) the rigid-flow kinematic moment of inertia ("kinematic" is defined as the value of the moment that does not include contributions from rotation-vibration-intrinsic interaction), and (ii) an interaction among the rotation, vibration, and intrinsic motions that is the product of the collective vibration displacements in the 1 and 3 directions (given by the second term on the L.H.S. in Eqs. (13) and (14)) and the product of the collective vibration displacement and intrinsic (radial) compression-dilation (given by the second and third terms on L.H.S. of Eq. (15) and the two terms on L.H.S. of Eq. (16)). As mentioned in Section 3, the interactions described in item (ii) above generate the fifth to tenth terms in the excitation energy in Eq. (19). These interactions and their consequential energy exchange between the oscillations in the 1 and 3 directions and between the vibration-rotation and intrinsic systems are responsible for reducing the rigid-flow kinematic moment of inertia to the measured value in the first $2^+$ excited state. This reduction arising from the interaction the rotation-vibration with a paired-nucleon intrinsic system parallels that achieved in the self-consistent cranking model by the addition of nuclear pairing correlations.

However, a change in the intrinsic-system configuration generated by the inclusion of a pairing interaction and the zero angular-momentum constraint in Eq. (1) would alter the interaction in item (ii) above modifying the behaviour of the moment of inertia with *J*. Indeed, the constraint in Eq. (1) may be replaced by $\langle \phi_{JK} | \vec{J} | \phi_{JK} \rangle = 0$ in the first approximation. This constraint is similar to the cranking-model type of constraint in the microscopic mean-field models [33], and would introduce a Coriolis interaction term into Eq. (15), generating excitation of paired particles into higher angular momentum orbitals. These changes in the intrinsic system would depend on *J* through the action of the term $\beta_1 \cdot \tilde{B}_3 + \beta_3 \cdot \tilde{B}_1$ in Eq. (15) because $\beta_1$ and $\beta_3$ depend on *J*, and hence would modify every term in Eq. (19) in each *J* state. These changes could give rise to back-bending before the band terminates at $J_c$ and the moment of inertia increases toward rigid-flow value.

## 6. Application of model to some nuclei

In this section, we present the predictions of the model for the ground-state rotational band in $^{8}_{4}Be$, $^{12}_{6}C$, $^{20}_{10}Ne$, $^{24}_{12}Mg$, $^{28}_{14}Si$, $^{162}_{66}Dy$, and $^{168}_{68}Er$. For each of the nuclei,: (i) we determine $\Sigma_k$ in Eq. (21) from its Nilsson's self-consistent deformed-oscillator particle configuration [54], and (ii) we adjust the values of the parameters $b_{int}$, $b_{v1}$, and $b_{v3}$ in Eqs. (21) and (24) to match as closely as possible the predicted and measured excitation energies and *B*(*E*2)'s (or the quadrupole moment) of the first excited $2^+$ state while ensuring that the rotational-band cutoff angular momentum $J_c$ is as high as possible when $\bar{a}_1$ is near unity.

Tables 1 and 2 show that the electric quadrupole moment $eQ_o$ and transition probability rates $B(E2)$'s are reasonably-well predicted when the measurement uncertainties are considered. The quadrupole moment is predicted to decrease with $J$ as in the $Sp(3,R)$ model [28,29], whereas it was predicted to monotonically increase with $J$ in the previous model [37,38], which ignored the interaction between the oscillations in the directions 1 and 3. Therefore, the interaction among the collective oscillations in the different spatial directions, and its consequential sharing of the momentum and energy among the oscillations, reduces the nuclear deformation (as discussed in Section 4). For this reason, the $B(E2)$'s are predicted to increase at low values of $J$ and decrease at higher values of $J$ as observed empirically, and as predicted by the $Sp(3,R)$ [28,29] and $SU(3)$ [39] models.

The results in Tables 1 and 2 show that the model predicts the excitation energy $\Delta E_J$ within +4% for $^8_4Be$, +2% and +9% for $^{12}_6C$, -13% and +48% for $^{20}_{10}Ne$, -14% and +18% for $^{24}_{12}Mg$, -28% and +26% for $^{28}_{14}Si$, -11% and +25% for $^{162}_{66}Dy$, and within 3% for $^{168}_{68}Er$. These results show that $\Delta E_J$ is reasonably-well predicted. However, in most cases $\Delta E_J$ is progressively over-predicted with $J$, and it appears to be caused by the interaction between the oscillations in the directions 1 and 3. The progressive over-prediction of $\Delta E_J$ with $J$ may be corrected by including in the model the dynamics of the oscillations in the direction 2, the zero angular-momentum constraint in Eq. (1), and possibly the non-axial terms ($\mathfrak{I}_{11}^{rig^{-1}} - \mathfrak{I}_{22}^{rig^{-1}}$) in Eq. (4). It is of interest to note that the systematic over-prediction of $\Delta E_J$ with $J$ appears to be also a feature of other models [29,30,34,57].

The predicted moment of inertia $\mathfrak{I}_J$, agrees closely with the measured moment, and is a factor of 2 or more smaller than the rigid-flow moment as expected. Therefore, the interaction between the oscillations in a pair of spatial directions and between the oscillation-rotation and intrinsic motions reduce the rigid-flow kinematic moment of inertia to the measured moment without using any pairing interaction. However, as a consequence of progressive over-prediction of $\Delta E_J$, the predicted moment of inertia $\mathfrak{I}_J$, defined in Eq. (26), decreases gradually with $J$ unlike the measured moment, which increases with $J$.

Tables 1 and 2 show that, for nuclei $^{162}_{66}Dy$, and $^{168}_{68}Er$, the predicted rotational-band cut-off angular momentum $J_c$ is realistically much higher than that predicted by the previous model [37,38], which neglected the interaction between the oscillations in the directions 1 and 3. It is expected that $J_c$ would be predicted even higher when the oscillations in the direction 2 is included in the model.

**Table 1. Predicted/measured excitation energy ($\Delta E_J$), Cut-off $J$ ($J_c$), moments of inertia ($\mathfrak{J}_J$), $eQ_o$, $B(E2)$**

| | $J$ | $\Delta E_J$ (MeV) model/exp | $2\mathfrak{J}_J/\hbar^2$ (MeV)$^{-1}$ model/exp | $2\mathfrak{J}_{rigflow}/\hbar^2$ (MeV)$^{-1}$ | Predicted $eQ_o$ /$B(E2)$ $e\,fm^2$ / $e^2\,fm^4$ | Measured $eQ_o$ /$B(E2)$ $e\,fm^2$ / $e^2\,fm^4$ |
|---|---|---|---|---|---|---|
| $^{8}_{4}Be$ | $2^+$ | 3.0/2.9 | 2.0/2.0 | 5.48 | 22/9 | 41 / 34±10.5 |
| | $4^+$ | 11.8/11.4* | 1.6/1.7 | 5.48 | 13/5 | Qo is for J=0 |
| | $J_c = 6^+$ | | | | (20/- at J=0) | (HF-BCS pre-diction for $2^+$ [58]) |
| $^{12}_{6}C$ | $2^+$ | 4.5/4.4 | 1.3/1.4 | 4.1 | -18/6 | -21±10.5 /11-99 Qo is for J=0 [59] |
| | $4^+$ | 15.3/14.1* | 1.3/1.5 | 4.1 | -27/7 | / 8.5 for $2^+$ [60,61] |
| | $J_c = 14^+$ | | | | (-19/- at J=0) | |
| $^{20}_{10}Ne$ | $2^+$ | 1.4/1.6 | 4.2/3.7 | 9.4 | 42/36 | 70±17.5$|_{J=0}$ /274-762 [59], 57±8 [61], 480±8 [62], |
| | $4^+$ | 4.9/4.3 | 4.1/5.4 | 9.4 | 40/46 | / 71±7 [61] |
| | $6^+$ | 10.2/8.8 | 4.1/4.9 | 9.4 | 38/46 | / 66±8 [61] |
| | $8^+$ | 17.9/12.0* | 4.0/9.5 | 9.4 | 35/40 | / 24±8 [61] |
| | $J_c = 10^+$ | | | | (42/- at J=0) | |
| $^{24}_{12}Mg$ | $2^+$ | 1.2/1.4 | 5.8/4.4 | 13.0 | 47/44 | 84±21$|_{J=0}$ /395-1097 [59] 119.3±25 [61], 425±29 [62] |
| | $4^+$ | 4.1/4.1 | 4.9/5.1 | 13.0 | 47/51 | / 95, +21,-16 [61] |
| | $6^+$ | 8.7/8.1 | 4.7/5.5 | 13.0 | 37/42 | / 140, +193, - 49 [61] |
| | $8^+$ | 15.6/13.2* | 4.4/5.9 | 13.0 | 27/25 | / 74, +148, - 29 [61] |
| | $J_c = 10^+$ | | | | (46/- at J=0) | |
| $^{28}_{14}Si$ | $2^+$ | 1.3/1.8 | 4.7/3.4 | 16.3 | -60/72 | $-38.5\pm21|_{J=0}$ /30.5-352.2 [59], 72±9 [61], 317±17 [62] |
| | $4^+$ | 4.6/4.6 | 4.3/4.9 | 16.3 | -72/147 | / 96±8 [61] |
| | $6^+$ | 10.7/8.5* | 3.6/5.6 | 16.3 | -90/254 | / 106±55 [61] |
| | $J_c = 6^+$ | | | | (-61/- at J=0) | |

**\*** No measured ground-state rotational-band energy level above this energy is reported in the Table of Isotopes and Nuclear Data Sheets.

**Table 2.** Predicted/measured excitation energy ($\Delta E_J$), Cut-off $J$ ($J_c$), moments of inertia ($\mathfrak{J}_J$), $eQ_o$, $B(E2)$

|  | $J$ | $\Delta E_J$ (MeV) model/exp | $2\mathfrak{J}_J/\hbar^2$ $(MeV)^{-1}$ model/exp | $2\mathfrak{J}_{rigflow}/\hbar^2$ $(MeV)^{-1}$ | Predicted $eQ_o$ /$B(E2)$ $e\,fm^2$ / w.u.[6] | Measured $eQ_o$ /$B(E2)$ $e\,fm^2$ / w.u. |
|---|---|---|---|---|---|---|
| $^{162}_{66}Dy$ | $2^+$ | 0.08/0.09 | 78/69 | 289 | 788/225 | -/ 204±3 [62] |
|  | $4^+$ | 0.26/0.27 | 78/78 | 289 | 787/321 | -/ 289±12 [62] |
|  | $6^+$ | 0.54/0.55 | 78/78 | 289 | 786/353 | -/ 301±17 [62] |
|  | $8^+$ | 0.92/0.92 | 78/81 | 289 | 786/368 | -/ 346±17 [62] |
|  | $10^+$ | 1.40/1.38 | 78/84 | 289 | 784/377 | -/ 350±23 [63] |
|  | $12^+$ | 1.99/1.90 | 78/88 | 289 | 783/383 | -/ 330±40 [63] |
|  | $14^+$ | 2.68/2.49 | 78/91 | 289 | 782/386 | -/ 330±40 [63] |
|  | $16^+$ | 3.47/3.14 | 78/96 | 289 | 780/388 | N/A |
|  | $18^+$ | 4.37/3.83 | 78/101 | 289 | 778/389 | N/A |
|  | $20^+$ | 5.37/4.58 | 78/105 | 289 | 776/389 | N/A |
|  | $22^+$ | 6.47/5.35 | 78/111 | 289 | 773/388 | N/A |
|  | $24^+$ $J_c=98^+$ | 7.67/6.15 | 78/117 | 289 | 771/387 (788/- at $J=0$) | N/A |
| $^{168}_{68}Er$ | $2^+$ | 0.08/0.08 | 80/75 | 308 | 718/189 w.u.[7] | -/ 213±4 [64] |
|  | $4^+$ | 0.25/0.26 | 80/76 | 308 | 717/266 w.u. | -/ 319±9 [64] |
|  | $6^+$ | 0.53/0.55 | 80/77 | 308 | 717/293 w.u. | -/ 424±18 [64] |
|  | $8^+$ | 0.91/0.93 | 80/79 | 308 | 716/306 w.u. | -/ 354±13 [64] |
|  | $10^+$ | 1.38/1.40 | 80/81 | 308 | 714/313 w.u. | -/ 308±13 [64] |
|  | $12^+$ | 1.96/1.94 | 79/84 | 308 | 713/317 w.u. | -/ 345±18 [64] |
|  | $14^+$ $J_c=94^+$ | 2.65/2.57* | 79/86 | 308 | 711/320 w.u. (718/- at $J=0$) | -/ 336 + 20, - 69 [64] |

\* No measured ground-state rotational-band energy level above this energy is reported in the Table of Isotopes and Nuclear Data Sheets.

## 7. Concluding remarks

An intuitively natural way to microscopically describe the collective rotation and vibration motion of deformed nuclei is to canonically transform the Hamiltonian to some collective angle, shape, and a set of intrinsic coordinates (as done for molecular rotation and vibration). However, over the last half a century, such an approach has had little success in producing practical results. The main reason for this outcome is that a suitable set of intrinsic coordinates, which would satisfy the transformation-related constraints imposed on the particle coordinates along the rotational axes, would destroy the simple single-particle (shell-model) picture of the intrinsic system. This would render difficult any calculation involving the intrinsic system and its interaction with the collective motions. (Some researchers have used rotating mean-field and group-theoretic approaches, cited in the introduction section, to circumvent this problem or deal with it directly using hyperspherical coordinates and group chains.)

---
[6] In weisskopt unit, 1 w.u. = 52.3 $e^2\,fm^4$.

[7] In weisskopt unit, 1 w.u. = 54.9 $e^2\,fm^4$.

In this article, we circumvent the difficulty associated with constrained intrinsic coordinates, discussed above, by canonically transforming, to the collective coordinates, the Schrodinger equation rather than the Hamiltonian, and by imposing the constraints on the wavefunction rather than on the body-fixed particle coordinates. The intrinsic wavefunction is required to be spinless (i.e., be a zero eigenstate of the angular momentum), and the rotational angles are chosen such that the transformed Schrodinger equation is that of a triaxial rigid-flow rotor plus an independent-particle constrained intrinsic component. The result is that, in the transformed Schrodinger equation, the Coriolis interaction term between the collective and intrinsic system vanishes. However, the Coriolis effect now appears in the intrinsic system in the form of zero angular momenum constraint on the intrinsic wavefunction. The rotor-plus-intrinsic Schrodinger equation is then transformed to three collective vibration coordinates, chosen to be the three principal-axes component of the rigid-flow moment of inertia tensor. The resulting partial differential equation is then reduced, using a constrained variational method, to four coupled, self-consistent, time-reversal invariant, cranking-type ordinary differential equations, and a self-consistency equation. The coupling self-consistency parameters appearing in these equations are the expectation values of collective vibration displacement and intrinsic compression-dilation operators. Finally, to simplify the solutions of equations, gain physical insight, and facilitate comparison with the remarkably successful phenomenological hydrodynamic Bohr-Davydov-Faessler-Greiner nuclear collective rotation-vibration model, we retain only the equations that are related to the collective and intrinsic motions in the direction 1 and 3.

For harmonic oscillator mean-field potentials, the governing equations are then solved in closed forms and the energy eigenvalue and self-consistency parameters are determined. The algebraic equations for these parameters are then solved interatively. In this way, we have reduced the solution of the nuclear Schrodinger equation to the solution of algebraic self-consistency equations.

The application of the model to the ground-state rotational band in the nuclei $^{8}_{4}Be$, $^{12}_{6}C$, $^{20}_{10}Ne$, $^{24}_{12}Mg$, $^{28}_{14}Si$, $^{162}_{66}Dy$, and $^{168}_{68}Er$ show that the electric quadrupole moment and hence the nuclear deformation decreases with $J$ because of the interaction the collective oscillations in the directions 1 and 3 and the resultant sharing of momentum and energy among the oscillations. The $B(E2)$ is predicted to increase at low values of $J$ and decrease at higher values of $J$, unlike that in the previous model. These results are consistent with those observed empirically and predicted by $SU(3)$ and $Sp(3,R)$ models.

The excitation energies are reasonable-well predicted, but they are progressively over-predicted with the angular momentum $J$, which seems to be a feature of other models as well. Because of the interaction between the oscillations in the 1 and 3 directions, the rotational-band cutoff angular momentum is predicted more realistically at high values of *J* in rare-earth nuclei. The predicted moment of inertia is close to that observed empirically. However, because of progressive overprediction of excitation energy, the predicted moment decreases gradually with *J* whereas the observed moment increases gradually with *J*. The model predicts that the interaction between the collective and intrinsic motions reduces the rigid-flow kinematic moment of inertia to the measured moment without using any pairing interaction, unlike that in the cranking model.

In a future study, we will address the progressive overprediction of the excitation energy, and hence the decrease in the moment of inertia with *J*. We will do so by generalizing the present model to include: (i) the oscillations in the 2 direction, (ii) the spinless intrinsic constraint and a possible connection of this constrain to quasi-particle rotation alignment and pair quenching, and (iii) non-axial terms in the rigid-rotor Schrodinger equation.